\documentclass[a4paper,12pt]{article}
\usepackage[utf8]{inputenc}

\usepackage{amsmath,amsfonts,amssymb,epsfig}
\usepackage{slashed}
\numberwithin{equation}{section}

\usepackage[titletoc]{appendix}
\usepackage{xfrac}

\usepackage{graphicx}
\usepackage{cancel}
\usepackage{cite}
\usepackage{color}
\usepackage{fancybox}
\usepackage{geometry}
\usepackage{lipsum}     
\usepackage{hyperref}

\usepackage{xspace}

\hypersetup{
    colorlinks,
    citecolor=black,
    filecolor=black,
    linkcolor=black,
    urlcolor=black
}
\def\s{\sigma}

\def\psib{\bar{\psi}}
\def\psih{{\psi_{\sfrac{1}{2}}}}
\def\psith{{\psi_{\sfrac{3}{2}}}}
\def\psibh{\bar{\psi}_{\sfrac{1}{2}}}

\def\d{\partial}

\def\ds{\slashed{\partial}}

\def\ks{\slashed{k}}
\def\qs{\slashed{q}}
\def\ps{\slashed{p}}
\def\us{\slashed{u}}

\def\g{\gamma}
\def\ptild{{\tilde{\pi}}}

\newcommand{\p}{\partial}

\newcommand{\Tr}{\mathop{\rm Tr}}

\def\ee{\varepsilon}

\newcommand{\half}{  \! {\sfrac{1}{2}}}
\newcommand{\thalf}{ \! \sfrac{3}{2}}

\def\eL{\epsilon_{ \! \! \! \! \textrm{ \ \tiny LV}}}

\def\L{\mathcal{L}}

\def\eps{\rho}

\def\PS{\mathbf{r}}
\def\PT{\mathbf{t}}
\def\P{\mathcal{P}}

\newcommand{\ov}[1]{\overline{#1}}

\newcommand{\beq}{\begin{equation}}
\newcommand{\eeq}{\end{equation}}
\newcommand{\beqs}{\begin{equation*}}
\newcommand{\eeqs}{\end{equation*}}

\newcommand{\bea}{\begin{eqnarray}}
\newcommand{\eea}{\end{eqnarray}}
\newcommand{\beas}{\begin{eqnarray*}}
\newcommand{\eeas}{\end{eqnarray*}}

\begin{document}

\begin{flushright}
\end{flushright}
\begin{center}

\vspace{1cm}
{\LARGE{\bf The Slow Gravitino}}

\vspace{1cm}

{\large{Karim Benakli$^{1,2,a}$,
\let\thefootnote\relax\footnote{$^a$kbenakli@lpthe.jussieu.fr}
Luc Darm\' e$^{1,2,b}$ \footnote{$^b$darme@lpthe.jussieu.fr} and Yaron
Oz$^{3,c}$ \footnote{$^c$yaronoz@post.tau.ac.il}}}
\vspace{0.7cm}

{\emph{$^1$ Sorbonne Universit\'es, UPMC Univ Paris 06,UMR 7589, LPTHE,\\
F-75005, Paris, France \\
$^2$ CNRS, UMR 7589, LPTHE, F-75005, Paris, France \\
$^3$ Raymond and Beverly Sackler
School of Physics and Astronomy \\
Tel-Aviv University, Ramat-Aviv 69978, Israel}}

\end{center}
\vspace{0.7cm}

\abstract{ 
When the supersymmetry breaking sector is a fluid background, Lorentz invariance is broken spontaneously.
 The super-Higgs mechanism  leads to a gravitino Lagrangian with Lorentz symmetry violating terms.  
 We analyse the resulting field equations and constraints.
 We identify the physical spin 3/2 and spin 1/2 helicity states, 
 derive their equations of motion and construct the propagator.
 The violation of Lorentz symmetry implies that the longitudinal mode has a non-relativistic dispersion relation, whose speed is
 lower than the speed of light.
 We briefly comment on possible implications to gravitino cosmology and phenomenology.

\newpage

\tableofcontents

\setcounter{footnote}{0}

 \vspace{1cm}

\section{Introduction and summary of the results}

The gravitino is massless
in Minkowski space-time when  supersymmetry is preserved, and acquires a mass when it is spontaneously
broken \cite{Fayet:1974jb,Deser:1977uq}. Supersymmetry breaking is necessary for a theory to 
be relevant to the real world, making its study being of much importance.

In the global supersymmetry limit, the breaking is associated with a non-vanishing vacuum expectation value of the Hamiltonian. 
 Lorentz invariance  implies that kinetic terms do not contribute. Promoting supersymmetry to be local, the contribution of the scalar potential to vacuum energy can be cancelled  by the presence of an additional  cosmological constant of opposite sign. The latter term breaks explicitly supersymmetry unless it comes along with a mass term for the gravitino, whose size is related to the scalar potential expectation value \cite{Deser:1977uq}. One ends up
finally with a massive gravitino in a space-time with a vanishing cosmological constant. 

 When supersymmetry is broken by the presence of a fluid, the kinetic energy is not necessarily vanishing allowing Lorentz symmetry violation.
 This scenario has been considered in  \cite{Benakli:2013ava}, and the aim of  this work is to study
this generalisation in more detail.

For a perfect fluid at rest, the breaking of boost invariance implies that the longitudinal and transverse modes
of the gravitino satisfy different dispersion relations. The longitudinal mode inherits its
dispersion relation  from that of the fluid goldstino \cite{Boyanovsky:1983tu,Hoyos:2012dh} and is non-relativistic.
The transverse mode
has the same mass  but has a relativistic dispersion relation \cite{Benakli:2013ava}. At high energies compared to the gravitino mass but well below the energy scale of the fluid, the gravitino interactions with matter fields are mainly via its longitudinal mode \cite{Fayet:1974jb,Casalbuoni:1988qd}.
The latter propagates at a speed  $\frac{p}{\rho}$, where $p$ and $\rho$ are
the the fluid pressure and energy density respectively. This speed is in general lower than the speed of light when the fluid stress-energy tensor satisfies
the appropriate energy conditions.
We therefore call it ``slow gravitino".

In \cite{Benakli:2013ava}, the gravitino mass term has been constructed in terms of the fluid pressure and
energy density for an arbitrary perfect fluid stress-energy tensor. 
However, the field equations were derived only for stress-energy tensors that are
constant or slowly varying only with time. In this work, we will generalise those
results  at quadratic order to a general space-time dependent perfect fluid.

There are various relevant scales and it is of importance to outline the 
approximations that will be made. First, there are the scales associated with the fluid.
The characteristic temperature scale ${\cal T}$ corresponds to the mean free path or correlation length of the
fluid microscopic degrees of freedom.
$L$ is the length scale over which the macroscopic fluid variables, i.e. the energy density,
pressure and fluid velocity, 
vary. In order for the hydrodynamics approximation to be valid one requires $\frac{1}{TL} \ll 1$.

Second, there is the energy scale of the propagating gravitino $E$ and the (reduced)  Planck mass $M_p$.
${\cal T}$ is also the supersymmetry breaking scale and we take ${\cal T} \ll M_p$.
We require the gravitino energy to be smaller than the supersymmetry breaking scale, $E\ll   {\cal T}$,  in order to be able
to keep only the goldstino interactions. Finally, we will require the gravitino wavelength to be smaller
than the fluid scale of changes in the macroscopic variables, $\frac{1}{EL} \ll 1$. This is required  in order to be able to consider
the gravitino as a localised particle with well defined helicity states.

Motivated by the implication of a vanishing mass super-traces  for the spectrum of phenomenological models,  
one usually assumes that supersymmetry is broken in a new sector of the theory, hidden or secluded, and then transmitted to our visible sector by mediator fields through gravitational or gauge interactions. In this work, the fluid under consideration will describe the hidden or secluded sector, and the scale ${\cal T}$ is
the supersymmetry breaking scale, which should not be confused with a temperature of our visible sector.

\subsection*{   }
\vskip -1cm
\subsubsection*{Fluid variables, scales and approximations}

We shall consider a gravitino  propagating in a
ideal fluid background specified by the energy density, pressure and velocity vector $u^\mu$ 
normalised as $u^\mu u_\mu = -1$. The fluid variables  are arbitrary slowly varying functions of the space-time
coordinates. The fluid stress-energy tensor reads
\begin{align}
 T_{\mu \nu} = \left[ p   \eta_{\mu \nu}+ (\eps  +p)u_{\mu} u_{\nu}\right] \ .
 \end{align}
We will use the equation of state 
\begin{equation}
w= \frac{p}{\rho} \ . \label{state}
\end{equation}
For 
$w\neq-1$ both supersymmetry and invariance under Lorentz boosts are spontaneously broken. The Lorentz invariant
cases with $F$ or $D$ term breaking correspond to a cosmological constant i.e. $w=-1$. 
 In the flat  space-time limit approximation, the super-Higgs mechanism leads to a gravitino mass \cite{Benakli:2013ava}
\begin{align}
\label{massgr}
m  &=  \frac{ \sqrt{ 3\eps }}{ 4 M_P} ~ |\frac{1}{3} - w| \ .
\end{align}
Note that, as discussed above, the gravitino mass is introduced in order to supersymmetrise the term that cancels  the contribution of
$\eta^{\mu \nu}T_{\mu
\nu}$ to the vacuum energy. Therefore, one expects  the mass to vanish when the trace of the stress-energy tensor vanish.
This is indeed the case in (\ref{massgr}). 

Using this explicit expression, imposing our last assumption  $\frac{1}{EL} \ll 1$ for gravitinos with energy of the order of their mass implies that
\begin{equation}
\label{neglectder}
 \frac{{\cal T}^2 L}{M_p} \gg 1 \ .
\end{equation}
In this approximation, we can neglect all derivatives of the fluid variables compared to the momentum or the mass of the gravitino.
We will work in this approximation.

In our Lagrangian, we will trade the fluid variables $\rho,p$ and use
\begin{align}
m, \qquad \eL \equiv 1+w , \qquad u^\mu \ ,
\end{align}
where $m$ is the mass (\ref{massgr}), and 
 $\eL$ is a dimensionless number that measures the
size of violation of Lorentz boost invariance. 
The Lorentz invariant solution corresponds to $w=-1$.

The fluid velocity is a time-like vector. 
We can define at every point in space-time two projectors $\PS$ and $\PT$ by
\begin{align}
\begin{split}
\label{PmunuDefFirst}
& \PS^{\mu \nu} \equiv \eta^{\mu \nu} + u^\mu u^\nu \\
 & \PT^{\mu \nu} \equiv (\mathbf{1} -  \PS)^{\mu \nu} = - u^{\mu} u^{\nu} \ .
\end{split}  
\end{align}
$\PT$ projects along $u^\mu$ , 
i.e. in the time-like direction defined by the fluid velocity, while $\PS$ projects on the vector space orthogonal to
$u^\mu$, i.e. on the spatial vector space defined by the fluid.

In general, the fluid velocity does not define a  foliation of space-time, unless the fluid is irrotational. 
However,  in the approximation~(\ref{neglectder}) the twist $v_{[\mu}\partial_{\nu}u_{\rho]}$ can be neglected. 
We can then
use  wave-functions   of the form $\psi^\mu \propto e^{i p^\mu x_\mu}$ with $p^{\mu}$ being
functions of the space-time coordinates, but their derivatives are neglected. We will call these wave functions plane-waves.
This will allow us to define helicitiy eigenstates and construct the corresponding propagator.

It is practical to work with the spatial and temporal components of
the gamma matrices $\g^\mu$ and the momentum $p^\mu$, defined via the projectors $\PS$ and $\PT$. They
are constructed  as
\begin{align}
\label{rmunuDefFirst}
r^\mu &= \PT^{\mu \nu} \g_\nu \ & k^\mu &= \PS^{\mu \nu} p_\nu \nonumber \\
 t^\mu &= \PS^{\mu \nu} \g_\nu \ & q^\mu &= \PT^{\mu \nu} p_\nu  \ .
\end{align}
$r^\mu$ and $t^\mu$ behave as $\g^i$ and $\g^0$.
They satisfy the relations $r^\mu r_\mu = -3$, $t^\mu t_\mu = -1$ and $t^\mu
r^\nu = -r^\nu 
t^\mu$.

\subsubsection*{Summary of the results}

The Lagrangian describing the gravitino field takes the form:
\begin{align}
\label{lagrangianMaster}
 \L =  \frac{1}{2}    \psib_\mu  \!  \! \left[ (\g^\mu \g^\nu + \eta^{\mu \nu})(-i\ds +  \! m )+ \!  i \g^\nu \d^\mu  \! 
 -i \g^\mu \d^\nu  \!  + \frac{3\eL m}{4-3\eL}  (r^\mu t^\nu + t^\mu r^\nu)
\right]  \! \psi_\nu \ .
\end{align}

In (\ref{lagrangianMaster}) one identifies the first term with the usual Rarita-Schwinger Lagrangian \cite{Rarita:1941mf}
and the term proportional to $\eL$ as the correction due to violation of Lorentz
invariance. This expression is not singular for $\eL=4/3$ because this
corresponds to the traceless energy-momentum tensor case where $m=0$.

For general fluid variables, we will decompose $\psi_\mu$ (see (\ref{decompositionfinal}))
into four spinors. They correspond to the helicity-$\frac{3}{2}$ states $\psi^\mu_{\thalf}$ and helicity-$\frac{1}{2}$  states $\psih$,
and two remaining spinors that are projected out by the constraints. Albeit $\psi^\mu_{\thalf}$ has a vector index which suggests that it
describes four spinors, these can be explicitly seen not to be independent in~(\ref{explicitdecompo})  
and in fact it  only contains two degrees of freedom corresponding to the helicities $\pm 3/2$ as expected from the overall counting.

In the approximation where the derivatives of the fluid variables can be neglected ~(\ref{neglectder}), 
the motion of the gravitino in a general fluid background can be presented in a similar way to its motion in a constant fluid background.
The difference being that we will use the time and space directions defined by the fluid instead of that of the laboratory frame.
 The field equations take the form
\begin{eqnarray}
\label{eoms}
 ( -i t^\rho \d_\rho   -i r^\rho\d_\rho +   m) \psi^\mu_{\thalf}   &=&~ 0  \ , \nonumber\\ 
(-i t^\rho \d_\rho +i wr^\rho \d_\rho + m ) \psih &=&~  0 \ .
  \end{eqnarray}
These equations of motion need to be supplemented with two constraints that project out the extra not-spin-3/2  degrees of freedom from $\psi^\mu$.
They read
\begin{equation}
-i \left[ r^\mu r^\nu + \PS^{\mu \nu}\right] \d_\mu \psi_\nu =  n r^\rho \psi_\rho \ , \label{C1}
\end{equation}
and 
\begin{equation}
(w r^\nu - t^\nu ) \psi_\nu = 0   \\    \\ \label{C2} \ ,
\end{equation}
where
\begin{equation}
 n ~\equiv~ \frac{m}{1-\frac{3}{4} \eL} \ .
\end{equation}
We find that the covariant propagator can be written as
\begin{align}
 G^{\mu \nu} &= \frac{\Pi_{~\thalf}^{\mu \nu}}{p^2 + m^2} + \frac{\Pi_{~\half}^{\mu
\nu}}{w^2 k^2 + q^2 + m^2  }  + \frac{3}{4} \eL \frac{\ks}{m k^2}(t^\mu k^\nu - 
k^\mu t^\nu) \ .
\end{align}
From this form one can see that the two parts corresponding to the helicity-3/2 and helicity-1/2 components of the spinor-vector 
have different poles, thus different dispersion relations. The quantities $\Pi_{~\thalf}^{\mu \nu}$ and 
$\Pi_{~\half}^{\mu \nu}$ are the corresponding polarisations.

\subsubsection*{Plan of the paper}

The plan of the paper is as follows. 
In section 2 we will derive the constraints. They  remove the  non-spin-$3/2$ states that are 
present in the original product of a vector and a spinor representations of the Lorentz group. 
Using the explicit form of these constraints will allow us to  decompose the gravitino field into its transverse and longitudinal modes.
  We will then derive the equations of motion for these modes and write the corresponding  Lagrangian.
  In section 3 we will derive the propagator and compare
it with the usual Rarita-Schwinger one \cite{VanNieuwenhuizen:1981ae}. 
In the discussion section we will outline 
 possible implications to gravitino cosmology and phenomenology.
Useful definitions, properties of projectors and  
details of calculations are give in the  appendices to ease the reading, while keeping the paper self-contained.

\section{Constraints and equations of motion}
\label{sec2}

Lorentz invariance is spontaneously broken in the fluid background, however
the notion of a state with spin quantum number (3/2 here) is still well defined.
 A spin-$3/2$ field can be built starting from a product of  spin-1/2 and spin-1 states and is
denoted as a fermion field carrying a vector index $\psi^\mu$. 
This is a reducible representation of the rotation group. Constraints have to be
used in order to extract the physical spin-3/2 degrees of freedom. 
We  present here the main aspects of such a construction  of a theory of spin-3/2 fields in
a fluid background. Additional details are provided in the appendices.

\subsection{The constraints equations}

We shall derive here two constraints that enable to reduce $\psi_\mu$ to its four degrees of freedom describing a massive spin-3/2 state.
 The equation of motion for $\psi_\mu$, obtained from the Lagrangian (\ref{lagrangianMaster}) is:
\begin{equation}
\label{Eomgeneral}
 K^{\mu \nu} \psi_\nu = 0 \ ,
 \end{equation}
with
\begin{equation}
\label{Eomgeneral2}
 K^{\mu \nu} ~\equiv~ (\g^\mu \g^\nu + \eta^{\mu \nu})(-i\ds + m )+i \g^\nu \d^\mu
 -i \g^\mu \d^\nu  + (n-m) (r^\mu t^\nu + t^\mu r^\nu)  \ .
 \end{equation}

In the Rarita-Schwinger case, a first constraint is  obtained by noting that the Lagrangian is linear in $\psi_0$,
 which therefore behaves as a Lagrange multiplier (see for example \cite{Deser:2000dz}). The  Euler-Lagrange equation for  
 $\psi^0$ gives   the time component of the equation of motion. This is used as a constraint as it contains no time-derivative.

In the fluid background case, we identify the time direction as the one given by the  fluid four-velocity $u^\mu$. 
We should therefore contract the equation of motion by $u^\mu$ 
to obtain the ``zeroth-component''. For calculation purposes we contract instead by $t^\mu = - \us u^\mu$ and use $\PT^{\mu \nu} + t^\mu t^\nu = 0$
 to obtain (\ref{C1}), which indeed does not contain  any time derivative (in the fluid frame).

Another constraint is obtained by contracting the equations of motion with ${\cal D}_\mu \equiv \d_\mu + E_\mu$ with 
\begin{equation}
E_\mu = - 
\frac{i}{2} n [ \g_\mu - \frac{3}{2}\eL t_\mu] \ .
\end{equation}
Using the form~(\ref{eomcovder}) of $K^{\mu \nu}$, the contraction gives
\begin{align}
\begin{split}
{\cal D}_\mu K^{\mu \nu} \psi_\nu  &= -i \epsilon^{\mu \nu \rho \sigma}  {\cal D}_\mu \g^5
\g_{\rho}  {\cal D}_\s \psi_\nu \\
 & = -i  \epsilon^{\mu \nu \rho \sigma}  \g^5 \g_{\rho}  E_\mu E_\s
  \psi_\nu \ ,
\end{split}
\end{align}
and after some algebra, this constraint reduces to
\begin{align}
T^{\mu \nu} \g_\mu \psi_\nu = 0 \ ,
\label{Tconstraint}
\end{align}
where replacing $T^{\mu \nu}$ by its expression and dividing by the energy density $\rho$ leads to~(\ref{C2}).

For a fluid at rest, (\ref{Tconstraint}) reads
\begin{equation}
 \g^0 \psi_0 = - ( 1- \eL ) \g^i \psi_i \ .
\end{equation}
When taking $\eL=0$, we recover the usual  Rarita-Schwinger constraint:
\begin{align}
 \g^\mu \psi_\mu = 0 \ .
\end{align}

To summarise, we have exhibited two constraints  (\ref{C1}) and (\ref{C2}) projecting out two  spin-1/2 states.
It is useful to note the similarities (up to derivatives terms) of our constraints with those obtained for the case of a gravitino in a Friedmann-Robertson-Walker (FRW) space in \cite{Kallosh:1999jj}.

\subsection{Identification of the spin-3/2 degrees of freedom}
\label{extractiondof}

We will now use the above constraints to identify the four degrees of freedom of our spin-3/2 state and write them as two transverse  (helicity-$3/2$) and two longitudinal (helicity-$1/2$) modes.

We first focus on the case where the fluid parameters are constant and work in the frame defined by the fluid background. In the last part of this section, we will  generalise the result for an arbitrary fluid where both translation and rotational invariance are lost but with the extra assumption~(\ref{neglectder}) implying that we can neglect derivatives in the hydrodynamics parameters.

In the constant fluid rest frame, the three-dimensional space is invariant under rotational and translation symmetries therefore both spin and helicity quantum numbers are well defined.
We start with representations of the Lorentz group but, as the boosts transformations are no more symmetries, we will work with representations of the rotations symmetry group i.e. spin representations. The left-handed spinor-vector representation of the Lorentz group (written as an $SU(2)_L \times SU(2)_R$ representation) can be decomposed into spin representations as
\begin{align}
\label{decompose}
(\frac{1}{2},\frac{1}{2}) \otimes (\frac{1}{2},0) =  \frac{1}{2} \oplus (1 \otimes \frac{1}{2}) = \frac{1}{2} \oplus \frac{1}{2} \oplus \frac{3}{2} \ .
\end{align}
The l.h.s. expresses $\psi^\mu$  as a tensorial product of a vector  times a spinor while  the last expression is a spin decomposition that can be written explicitly as a linear combination of normalised  spin eigenstates. Using the Clebsch-Gordon decomposition this leads to:
\bea
\label{explicitdecompo}
\psi^\mu ~= &&  {\epsilon'_0}^\mu ~\tilde{a}_1\xi'_- +  {\epsilon'_0}^\mu ~a_1 \xi'_+  \nonumber \\
& + &   \frac{1}{\sqrt{3}} {\epsilon_0}^\mu(   \tilde{a}_2 \xi_- - a_2 \xi_+) + \sqrt{\frac{{2}}{{3}} }~ ( \epsilon_+^\mu ~a_{2} \xi_- - \epsilon_-^\mu ~\tilde{a}_{2} \xi_+) \nonumber \\
& + &  \sqrt{\frac{{2}}{{3}} } ~ {\epsilon_0}^\mu( \tilde{a}_{3} \xi_- + a_{3} \xi_+) + \frac{1}{\sqrt{3}}  ( \epsilon_+^\mu 
~a_{3} \xi_- + \epsilon_-^\mu ~\tilde{a}_{3} \xi_+)  \nonumber  \\
& + &   \epsilon_-^\mu ~\tilde{a}_4 \xi_-   +  \epsilon_+^\mu  ~a_4 \xi_+  \ .
\eea
The two first lines correspond the extra spin-1/2 representations that need to be projected out of the spectrum. The third line is the helicity $\pm 1/2$ part, while the last line is the helicity $\pm 3/2$ part of the spin-3/2 representation of interest. 
The coefficients $a_{i}$ and $\tilde{a}_i$ parametrise the decomposition as function of the product of polarisation vectors $\epsilon^\mu_i$ and spinors $\xi_i$ and $\xi'_i$. The indices of the latter vectors and spinors give their respective helicity eigenvalues in a self-explanatory way. The physical degrees of freedom must  satisfy both constraints~(\ref{C1}) and~(\ref{C2}).

It is convenient to introduce a spinor that describes the longitudinal  degrees of freedom  of our spin-3/2 field.  This is achieved by defining $\psih$ from our explicit construction~(\ref{explicitdecompo}) as

\begin{align}
\psih ~ \equiv ~ \frac{n}{w |p|} (a_{3} \xi_+ + \tilde{a}_{3} \xi_-) \ ,
\end{align}
where $|p| = \sqrt{-p^2}$. The overall coefficient ensures that $\psih$ has the canonically normalised kinetic term for a Majorana spinor. We shall show below that this can be written as 
\begin{align}
\label{psihrest}
\psih = \sqrt{\frac{3}{2}} \frac{ n}{ k} \g_0 \g^i \psi_i \ .
\end{align}

We consider now the case of a generic fluid under the assumption~(\ref{neglectder}). It is possible to find the corresponding form of $\psih$ either from the requirement that the constraints are satisfied or through an explicit construction. We shall use the former.

In order to identify among the  physical spin-1/2, we consider an operator $\Pi_\mu$  that satisfies the constraints~(\ref{C1}) and~(\ref{C2}) written as:
\begin{align}
\label{defC}
\begin{split} C_1^\mu \Pi_\mu &= 0 \\ C_2^\mu \Pi_\mu &= 0 \end{split} & \textrm{  with } &  \begin{split} C_1^\mu &=  w r^\mu - t^\mu \\ C_2^\mu  &= (k^\mu + \ks r^\mu) - n r^\mu \ . \end{split}
\end{align}
Such $\Pi_\mu$ is given by
\begin{align}
\label{pidef}
\Pi^\mu = (r^\mu - 3 \frac{\ks k^\mu}{k^2}) - \frac{2}{n} k^\mu - \frac{2w}{n} \ks t^\mu \ ,
\end{align}
and we will also define a conjugate operator as $ \bar{\Pi}^\mu$ (note the change of sign of the last term)
\begin{align}
\label{ptilddef}
\bar{\Pi}^\mu = (r^\mu - 3 \frac{\ks k^\mu}{k^2}) - \frac{2}{n} k^\mu + \frac{2w}{n} \ks t^\mu \ ,
\end{align}
Solutions of the constraints  can then be obtained through projection by the operator $P_{\half}$  ( note that $P_{\half} P_{\half} = P_{\half}$ )   defined by 
\begin{align}
 P_{\half}^{\mu \nu} = \Pi^\mu (\bar{\Pi}^\rho \Pi_\rho)^{-1} \bar{\Pi}^\nu \ .
\end{align}
Using the constraint~(\ref{C2}) we can write
\begin{align}
\psih^\mu ~\equiv~ P_{\half}^{\mu \nu} \psi_\nu= \frac{\Pi^\mu}{2} \frac{n \ks}{k^2} ~ r^\rho \psi_\rho \  ,
\end{align}
which define the helicity-1/2 part $\psih^\mu$ of $\psi^\mu $. We can write a corresponding spinor with the same degrees of freedom, a canonically normalised kinetic term in the Lagrangian, but without vector indices.  It is obtained by contraction with $u_\mu$:
\begin{align}
\label{psi12}
\psih = - \sqrt{\frac{3}{2}} \frac{n}{w k} u_\mu P_{\half}^{\mu \nu} \psi_\nu = \sqrt{\frac{3}{2}} \frac{n}{k} \slashed{u} ~ r^\rho \psi_\rho \ ,
\end{align}
and describes the longitudinal modes of the gravitino.  Note that in the rest frame we recover (\ref{psihrest}).

The helicity $\pm 3/2$ degrees of freedom can be identified as the remaining modes of $\psi^\mu$ after  removing all three independent spin-1/2 states of the vector-spinor state. Such spinors can be constructed by applying on $\psi^\mu$ the three orthogonal projectors $\P_{ii}$:
\begin{equation}
{\tilde \P}_{ii}^{\mu \nu} = \displaystyle \frac{\ptild_i^\mu
\ptild_i^\nu}{\ptild_i^2}  \ ,
\end{equation}
where $\ptild^\mu_1$, $\ptild^\mu_2$ and $\ptild^\mu_3$ are orthogonal operators defined such that $\ptild_i^\mu \ptild_{j,\mu} =0$ if $i \neq j$ by:
\begin{align}
\begin{split}
\ptild_1^\mu &= t^\mu  \\
\ptild_2^\mu &= r^\mu \\
\ptild_3^\mu &= r^\mu - 3 \frac{\slashed k k^\mu}{k^2}   \ .
\end{split}
\end{align}
Note that $\Pi_\mu$ can be expressed as a linear combination of these. The corresponding projector $\P_{\thalf}$ is given by
\begin{equation}
\label{P3/2}
\P_{\thalf}^{\mu \nu} =
\eta^{\mu \nu} - {\tilde \P}_{33}^{\mu \nu} -{\tilde \P}_{22}^{\mu \nu} -{\tilde \P}_{11}^{\mu \nu} \ ,
\end{equation}
and $\psi_{\thalf}^\mu ~\equiv~ P_{\thalf}^{\mu \nu} \psi_\nu$ corresponds to the transverse degrees of freedom.
This can be expressed as
\begin{equation}
\label{psi32}
\psi^\mu_{\thalf} =   \psi^\mu  + \frac{1}{3} r^\mu r^\nu \psi_\nu + t^\mu t^\nu \psi_\nu
+ \frac{1}{6} (r^\mu - 3 \frac{\ks k^\mu}{k^2})(r^\nu - 3\frac{\ks k^\nu}{k^2})
\psi_\nu \ .
\end{equation}
Using the fact that   $r_\mu \psi^\mu_{\thalf} = t_\mu \psi^\mu_{\thalf} = k_\mu \psi^\mu_{\thalf} =0$, it is easy to check that $\psi^\mu_{\thalf} $ satisfies    the constraints~(\ref{C1}) and~(\ref{C2}) and also that $ P_{\half}^{\mu \nu} \psi_{~\thalf ~\nu} =0$. We  chose to keep the vector indices to remind of its spin-3/2 nature.

To summarise, in the space of solutions of the constraints ~(\ref{C1}) and~(\ref{C2}), we have the decomposition 
\begin{equation}
\label{decompositionfinal}
\psi^\mu = \psi_{\thalf}^\mu + \Pi^\mu \frac{\ks \us}{\sqrt{6} k} \psih  = \P_{\thalf}^{\mu \nu} \psi_\nu + \P_{\half}^{\mu \nu} \psi_\nu \ ,
\end{equation}
where the two terms corresponds to the transverse and longitudinal modes of the spin-3/2 field. 

While for the fluid at rest the helicity was defined as the projection on  the globally defined direction corresponding to the space component of the particle momentum, the definition is more involved in the case of fluid not at rest as, in general, plane waves are no longer solutions of the equation of motion. However, helicity can be defined  under the assumption (\ref{neglectder}).

In the rest frame the helicity operator is defined as:
\begin{align}
\mathcal{S} ~=~ \frac{1}{2} \epsilon_{ijk} S^{ij} \d^k \ ,
\end{align}
where the $S^{\nu \rho}$ are the Lorentz generators for the spin-3/2 representation. This generalises to 
\begin{align}
\label{helicity}
\mathcal{S}  ~\equiv~ \frac{1}{2} u^\mu \epsilon_{\mu \nu \rho \sigma} S^{\nu \rho} \PS^{\s \g} \d_\g \ ,
\end{align}
as $\PS^{\s \g} \d_\g$ reduces to the space derivatives in the fluid rest frame. In a space-time varying fluid but with the assumption~(\ref{neglectder}), locally we can  treat the eigenstates wave-functions as plane waves. As a consequence, the above decomposition can be carried over and locally $\psih^\mu$ and $\psith^\mu$ appear as helicity eigenstates with eigenvalues respectively $\pm \frac{1}{2}$  and $\pm \frac{3}{2}$ .

\subsection{The equations of motion}

In this section we will derive the equation of motions~(\ref{eoms}) for the fields $\psih$ and
$\psi^\mu_{3/2}$ corresponding to the longitudinal and transverse modes  of the
massive spin-$3/2$ state as defined above. We neglect the derivatives in the fluid parameters according to~(\ref{neglectder}).

The equation of motion for $\psi^\mu$ derived from the Lagrangian are given in~(\ref{Eomgeneral}).
 In order to extract those for the $\psih$ and
$\psi^\mu_{3/2}$,  it is useful to use the identity
 \begin{align}
 \label{Eomintermed}
 [i  \g^\nu \d^\mu - i \g^\mu \d^\nu] \psi_\nu = [t^\mu r^\nu + r^\mu t^\nu] (i \ds - n)\psi_\nu - \bar{C}_2^\mu \g^\nu \psi_\nu \ ,
\end{align}
where $ \bar{C}_2^\mu$ is given by:
\begin{equation}
 \bar{C}_2^\mu = k^\mu + r^\mu \ks - n r^\mu \ .
\end{equation}
Plugging (\ref{Eomintermed}) in (\ref{Eomgeneral}) leads to
\begin{align}
\label{EomC2}
\left[ (r^\mu r^\nu + \PS^{\mu \nu})(-i\ds + m )+ \bar{C}_2^\mu \g^\nu \right] \psi_\nu = 0 \ , 
\end{align}
which will be used to derive the equations of motion for both helicities 1/2 and 3/2.

We first focus on the helicity-1/2 degrees of freedom.
We can get rid of the term proportional to $ \bar{C}_2^\mu$  by contracting~(\ref{EomC2}) with $\bar{\Pi}^\mu$ defined in~(\ref{ptilddef}). Two parts of the equations  are obtained through splitting the derivative in the l.h.s. to the time-like and space-like parts. We consider plane waves solutions.  A bit of algebra allows to rewrite  the space-like part, along with the mass term, as 
\begin{align}
  \bar{\Pi}_\mu  (r^\mu r^\nu + \PS^{\mu \nu})(\ks + m ) \psi_\nu = \sqrt{6} \frac{\ks \us}{k} ( v \ks - m) \psih \ .
\end{align}

On the other side, the time-like part, using the decomposition~(\ref{decompositionfinal}) , can be expressed as 
\begin{align}
\bar{\Pi}_\mu (r^\mu r^\nu + \PS^{\mu \nu}) \qs \psi_\nu = &  - \sqrt{6} \frac{\ks \us}{k} \qs \psih  \ .
\end{align}
Putting back both parts together leads to the equation of motion for the longitudinal mode:
\begin{align}
 \label{Eom12}
( -i t^\rho \d_\rho +  i w r^\rho \d_\rho + im ) \psih = 0 \ .
\end{align}

In order to derive the equation of motion for the transverse degrees of freedom, we act on~(\ref{EomC2}) with the operator ${\P_{\thalf}}$ to obtain
\begin{align}
( \qs + \ks + m ) \psith^\mu   = 0 \ ,
\end{align}
which can be written:
\begin{align}
( -i t^\rho \d_\rho - i r^\rho \d_\rho\ +  m ) \psith^\mu =   0 \ .
\end{align}

An interesting consequence of these equations of motion is that helicity-1/2 and 3/2 cannot be on-shell simultaneously when $\eL \neq 0$.

The equations of motion~(\ref{eoms}) derived above can also be obtained from the Lagrangian
\begin{align}
\label{first_order_lagrangian}
\L ~=~ \frac{1}{2} \bar{\psi}_{\ \thalf \ \mu} ( -i \g^\rho \d_\rho   +  m ) \psith^\mu + \frac{1}{2}  \psibh (- i t^\rho \d_\rho +   i w r^\rho \d_\rho +  m ) \psih \ , 
\end{align}
where we verify that the factors in the definition of $\psih$ were necessary for obtaining a canonically normalised kinetic term.  The two spinors have obviously different dispersion relations. The hermiticity  of the Lagrangian requires
that derivatives of the fluid parameters are neglected as we assumed in (\ref{neglectder}).

\section{The covariant spin--3/2 propagator}

In this section we aim at calculating the propagator of $\psi_\mu$ without using
the on-shell constraints. In order to work in the Fourier space and obtain an explicit form of the propagator, 
we need to consider a constant fluid background or more generally work in the approximation (\ref{neglectder}) where on can neglect derivatives in the fluid variables.

The  calculations are long and tedious. We review here  the main lines and present many details in the  Appendix~\ref{AppPropagator}. 
Our strategy will consist of writing the Lagrangian in a basis of projectors
adapted to the degrees of freedom of our problem. It is convenient to use as  basis the 
$\pi_i^\mu$, $(i=1,2,3)$ defined by
\begin{align}
\begin{split}
 \pi_1^\mu &= p^\mu  \\
 \pi_2^\mu &= \ks (t^\mu - \frac{\slashed q p^\mu}{p^2}) \\
 \pi_3^\mu &= r^\mu - \frac{\slashed k k^\mu}{k^2}   \ .
\end{split}
\end{align}
The Lagrangian involves the quantities  $p^\mu, \g^\mu , r^\mu$ and  $t^\mu$
that can be  expressed as
\begin{align}
 \begin{split}
p^\mu  &=   \pi_1^\mu \\
 \g^\mu &=   \frac{\ps}{p^2} \pi_1^\mu - \frac{\ps}{k^2} \pi_2^\mu + \pi_3^\mu
\\
 r^\mu &=   \frac{\ks}{p^2} \pi_1^\mu - \frac{\qs}{k^2} \pi_2^\mu + \pi_3^\mu \\
 t^\mu &=   \frac{\qs}{p^2} \pi_1^\mu - \frac{\ks}{k^2} \pi_2^\mu  \ .
\end{split}
\end{align}
We can define three projectors $\P_{i,i}$ as
\begin{equation}
 \P_{i,i}^{\mu \nu} = \frac{\pi_i^\mu \pi_i^\nu}{\pi_i^2} \ ,
\end{equation}
and supplement them by nilpotent operators $\P_{i,j}$ with $i \neq j$ defined by 
\begin{equation}
 \P_{i,j}^{\mu \nu} = \frac{\pi_i^\mu \pi_j^\nu}{\pi_i^2 \pi_{j}^2} \ ,
\end{equation}
where  $\pi_i^2 = \pi_i^\mu \pi_{i, \mu} $. We then define similarly $\P_{\thalf}$ orthogonal to all the other projectors by
\begin{align}
\P_{\thalf}^{\mu \nu}= \eta^{\mu \nu} - \P_{33}^{\mu \nu} - \P_{22}^{\mu \nu} - \P_{11}^{\mu \nu} \ ,
\end{align}
and we have checked this is the same projector as the one defined in equation  (\ref{P3/2}).
The normalisation has been chosen to be similar for $\P_{i,j}$ and
$\P_{j,i}$, a choice which helps making all the expression explicitly symmetric
but slightly complicates the algebra of these operators. This allows to write the quadratic operator $K^{\mu
\nu}$ as
\begin{align}
 K^{\mu \nu} = (\ps + m ) (\P_{\thalf}^{\mu \nu} - \P_{3 3}^{\mu \nu}) - U
(\P_{13}^{\mu \nu} - \P_{31}^{\mu \nu}) + V (\P_{23}^{\mu \nu} - \P_{32}^{\mu
\nu}) - W (\P_{12}^{\mu \nu} - 
\P_{21}^{\mu \nu}) \ ,
\end{align}
with
\begin{align}
&W = n k^2 \\[0.8em] 
&U = 2 (\ks m + \qs n) \\
&V = 2 \frac{k^2}{p^2} (p^2 - m \qs - n\ks)  \ .  
\end{align}
The propagator $G^{\mu \nu}$ is decomposed in the projector basis as 
\begin{align}
G^{\mu \nu} = \frac{ m  - \ps}{m^2 + p^2} \P_{\thalf}^{\mu \nu} + & A ~ \P_{1
1}^{\mu \nu} + B ~ \P_{2 2}^{\mu \nu} +  C ~\P_{3 3}^{\mu \nu} + D ~\P_{13}^{\mu
\nu}  + D' ~\P_{31}^
{\mu \nu}  \\ a
& + E ~\P_{23}^{\mu \nu} + E' ~\P_{32}^{\mu \nu} + F ~\P_{12}^{\mu \nu}  + F'
~\P_{21}^{\mu \nu} \ ,
\end{align}
where $A = A_1 + \ks A_2 + \qs A_3$ and similarly for all other coefficients. We
look for a solution of the equation defining the propagator in momentum space:
\begin{equation}
 K_{\mu \rho} G^{\rho \nu} = \eta_\mu^{~\nu}
\end{equation}
The result can be expressed as:
\begin{align}
 G^{\mu \nu} &= \frac{\Pi_{~\thalf}^{\mu \nu}}{p^2 + m^2} + \frac{\Pi_{~\half}^{\mu
\nu}}{w^2 k^2 + q^2 + m^2  }  + \frac{3}{4} \eL \frac{\ks}{m k^2}(t^\mu k^\nu - 
k^\mu t^\nu) \ .
\end{align}
where the two polarisations take the form 
\begin{align}
\Pi_{~\thalf}^{\mu \nu} = & (m - \ps) \P_{\thalf}^{\mu \nu} \ ,
\end{align}
and
\begin{align}
\begin{split}
\Pi_{~\half}^{\mu \nu} =&  - \frac{2}{3}\,  \,  \Lambda^\mu \,  \,  (\ps  - \eL \ks + m)\,  \,  \Lambda^\nu  \ , 
\end{split} 
\end{align}
where
\begin{equation}
\Lambda^\mu = \g^\mu + \frac{p^\mu}{n} - \frac{3}{2} (r^\mu - \frac{\slashed k k^\mu}{k^2}) - \frac{3}{4} \eL t^\mu  \ ,
\end{equation}
Note that we recover again that the part corresponding to the spin-1/2 components of the spinor-vector
has a pole for $m^2 + w^2 k^2 + q^2 = 0$ due to a different dispersion relation. A crucial observation is that the nominator of the helicity-1/2 poles indeed projects on the physical degrees of freedom. More precisely one can show that 
\begin{align}
\Pi_{~\half}^{\mu \nu}  \psi_\mu = \Pi_{~\half}^{\mu \nu}  \psi_{~\half~\mu} \ .
\end{align}

The modification of the Rarita-Schwinger propagator due to Lorentz symmetry
breaking appears both in the spin-3/2 and 1/2 contributions. When $\eL = 0$, we
recover the usual Rarita-Schwinger formula \cite{VanNieuwenhuizen:1981ae}. 

We finally consider the limit of high momentum where we have the hierarchy
\begin{align}
m ~ \ll ~ |p| ~\ll ~{\cal T} \ ,
\end{align}
the propagator then simplifies to
\begin{align}
 G^{\mu \nu} &\rightarrow  - \P_{\thalf}^{\mu \nu} \frac{ \ps}{p^2}   - \frac{2}{3}\frac{p^\mu p^
\nu}{n^2} \frac{\qs - w \ks}{q^2 + w^2 k^2 } \ .
\end{align}

\section {Discussion}
Two main tasks have been achieved in this paper. First, we have studied the propagation of a gravitino in a fluid background responsible of the breaking of supersymmetry. We exhibited a form of the Lagrangian where the Lorentz symmetry breaking terms are isolated explicitly making  it easy to compare to the well known 
Rarita-Schwinger case. The spin-$3/2$ propagating degrees of freedom have been identified and the constraints needed to remove
the other field components have been  derived.  The splitting of the degrees of freedom has been carried on 
in the case of a generic fluid where where the energy density, the pressure and the fluid velocity
are slowly varying functions of space-time coordinates so that we can neglect fluid derivatives. This generalises the results of \cite{Benakli:2013ava}. 

A second main result is the corresponding explicit formula for the propagator. This is a prerequisite if one wants to compute amplitudes in covariant form where the spin-$3/2$ is involved.  As an aside, we have introduced a set of projectors, quantities and notations which seems to us  helpful to carry on similar computations.

We end the paper by some comments  on possible phenomenological implications.  First, note that if the fluid background responsible of the breaking of 
supersymmetry is to be identified with our visible universe it has to be at the very early stages of its evolution. At later time, temperature is too low to explain the 
experimentally required size of supersymmetry breaking while the induced Lorentz violation will be too large. At early time, our discussion should be extended to 
curved background which should be straightforward.

We therefore consider, as stated in the introduction, that the fluid under discussion  describes a hidden sector gravitationally coupled to our universe. An important question is then if in such a case the induced Lorentz violation is small enough to be allowed today, or whether
 it should be restricted to early cosmological times. A definite answer to this question requires computing this effect in a given model.
 We can nevertheless discuss it qualitatively. Because the hidden sector fluid is only coupled gravitationally to the visible sector, the induced Lorentz violation for Standard Model fields must go to zero when $M_P$ is taken to  be infinite. We parametrise the shift in the effective speed of light in visible sector dispersion relations as  $(\frac{{\cal T}^2}{M_P^2})^\alpha \eL \sim \frac{{m}^\alpha}{M_P^\alpha} \eL$. This is constrained to not exceed about $\frac{\Delta c}{c} \approx 10^{-22}$ \cite{Coleman:1998ti}. It translates to a bound on a combination of 
$\eL$ and $m$.  Taking as an example a reference value of $m =1$ TeV shows that an $\eL \sim {\cal O} (1)$ is allowed for a value of $\alpha = 3/2$, which is not unreasonable. The Planck suppression and supersymmetry could then account for a small Lorentz violation in our visible world.

When discussing gravitino interactions, one can consider two energy regimes. For $E < m$ all modes of the gravitino couple with similar Planck suppressed strength while for  $E > m$ the coupling of the helicity 1/2 mode is enhanced,  only suppressed   by ${\cal T}^{-2}$. This is important for early universe history as one usually assumes that  gravitinos can be very energetic and their interactions approximated by those of the longitudinal mode. However, the violation of Lorentz invariance has also an important consequence for this mode as it    has been pointed out in \cite{Coleman:1998ti}. At high energies, when the mass terms can be neglected, the longitudinal mode of the gravitino could be stable because it can decay only to particles with higher effective speed of light. The usual analysis of the effects of gravitino in cosmology would need to be reconsidered.

Finally, we would like to stress that while our work often refers to the gravitino, the Lagrangian might also describe the propagation of other (composite)
spin-3/2 state in a Lorentz symmetry violating background.

\section*{Acknowledgments}
Y.O. would like to thank the LPTHE and ILP for their hospitality. This work  was supported in part by French state funds
managed by the ANR Within the Investissements d'Avenir programme under reference
ANR-11-IDEX-0004-02 and in part by the ERC Higgs@LHC.
The work of Y.O.  is supported in part by the I-CORE program of Planning and Budgeting Committee (grant number 1937/12), and by the US-Israel Binational Science Foundation.

\newpage

\appendix
\addcontentsline{toc}{section}{Appendices}
\section*{Appendices}

\section{Super-Higgs mechanism in a fluid}
\label{AppSuperHiggs}

We start by a brief review of some of the results obtained in 
\cite{Benakli:2013ava}. These will be recast  in the form that will be
used for the computations described in the paper. 

We consider a supersymmetric fluid with stress-energy tensor $T^{\mu\nu}$.  The 
presence of a goldstino  
associated with the spontaneous breaking of supersymmetry is expected from the
Ward-Takahashi identity. This Majorana state has been named phonino
 \cite{Boyanovsky:1983tu} and satisfies the equation of motion:
\beq
T^{\mu\nu}\g_{\mu}\partial_{\nu}{G} = 0 \ .
\label{Tfeqgold}
\eeq
which can be derived  from the Lagrangian where we neglect derivatives of fluid variables:
\beq
{\cal L}_{G} = -\frac{i}{{2\cal T}^{4}}T^{\mu\nu}
\bar{G} \g_{\mu}\partial_{\nu}G  \  .
\label{eqGT}
\eeq
Here, ${\cal T} \neq 0$ can be chosen to be ${\cal T} = | \Tr \,  T^{\mu\nu} |^{\frac{1}{4}}$ or ${\cal T} = | \det\, T^{\mu\nu}  |^{\frac{1}{16}}$ and has
dimension of mass.
Note that for $T^{\mu\nu} = -|F|^2 \eta^{\mu\nu}$, the Lagrangian (\ref{eqGT})
reduces to that of the usual goldstino.

Within the hypothesis of working in the flat Minkowski space-time limit, we will
 study the Lagrangian describing the
system phonino-gravitino at the quadratic order and keep the lowest order of an
expansion in powers of 
the dimensionless parameter $\frac {\cal T}{M_p}$. It reads:
\begin{align}
\label{finalL}
\L ~=~& - \frac{i}{2} \epsilon^{\mu\nu\rho\sigma} \psib_\mu \g^5 \g_\nu \d_\rho
\psi_\s - \frac{1}{4} \epsilon^{\mu\nu\rho\s} n_{\s \lambda} \psib_\mu \g^5
\g_\rho \g^\lambda \psi_\nu 
 - \frac{i}{\sqrt{2}}  \frac{{\cal T}^2}{M_P }  \frac{T^{\mu\nu}}{ {\cal T}^4}  
 \psib_\mu \g_\nu G  \nonumber \\
 &+  i \frac{T^{\mu\nu}}{ 2 {\cal T}^4}  \bar G \g_\mu \d_\nu G + \frac{1}{4}
\frac{T^{\mu\nu} n_{\mu\nu}} {{\cal T}^4} \bar G  G \ .
\end{align}
This Lagrangian is invariant under the supersymmetry transformations with
Lorentz
violating coefficients
\begin{align}
 \delta G  &= \sqrt{2} {\cal T}^2 \ee \ ,
\nonumber \\
\delta \psi_{\mu} &= - M_P ( 2 \p_\mu \ee  + i  n_{\mu\nu} \g^\nu \bar \ee
) \ , 
\end{align}
if $n_{\mu\nu}$ satisfies: 
\beq
\label{general mass}
-\frac{1}{2} \epsilon^{\mu\nu\sigma\rho} \epsilon_{\rho}^{~\lambda\gamma\kappa}
n_{\nu\lambda} n_{\sigma\gamma} = 
\frac{T^{\mu\kappa}}{M_P^2} \ .
\eeq

In the unitary gauge,  $G$ is set to zero through the supersymmetry
transformation:
\begin{align}
\psi_{\mu } \rightarrow \psi_{\mu }  + 
\frac{\sqrt{2}M_P}{{\cal T}^2} \d_\mu G + i  
\frac{M_P}{\sqrt{2}{\cal T}^2}n_{\mu\nu} 
\g^\nu \bar G  \,.
\end{align}

For multiple fluids without interactions among them, one should only replace
$T^{\mu \nu}$ by the sum $\sum_{i} T_i^{\mu \nu}$.

The above expression of the gravitino quadratic term  did not assume a perfect fluid or
another specific form, but rests on the assumption of existence of a goldstino
with the appropriate dispersion relation. An explicit dependence of the mass on
the fluid parameters can be obtained for a perfect fluid with a four-velocity $u^{\mu}$ such that $u_{\mu} u^{\mu} = -1$
\begin{align}
 \label{Tperfect}
 T_{\mu \nu} = \eps  \left[ w \eta_{\mu \nu}+ (1+w)u_{\mu} u_{\nu}\right] \ .
 \end{align}
This expectation value of the stress-energy tensor  breaks spontaneously both
supersymmetry and Lorentz symmetry. 

After choosing the unitary gauge, the Lagrangian~(\ref{finalL}) takes the form~(\ref{lagrangianMaster})
\begin{align}
\L ~=~ \frac{1}{2}\psib_\mu K^{\mu \nu}  \psi_\nu \ ,
\end{align}
where  $K^{\mu \nu}$ can be split into a kinetic and mass term
\begin{align}  
K^{\mu \nu} &= -i(\g^\mu \g^\nu + \eta^{\mu \nu})\ds +i \g^\nu \d^\mu  -i \g^\mu \d^\nu  + K_m^{\mu \nu} \ , 
\end{align}
with
\begin{align}  
 K_m^{\mu \nu} &=   m \left[ \g^\mu \g^\nu + \eta^{\mu \nu} + \frac{3\eL }{4-3\eL}  (r^\mu t^\nu + t^\mu r^\nu) \right] \ .
\end{align}
It will be useful to note that it can also be written as
\begin{align}
\label{eomcovder}
K^{\mu \nu} &= -i\epsilon^{\mu \nu \rho \sigma}  \g^5 \g_{\rho} {\cal D}_\s     \  ,
\end{align}
where we use a  derivative operator defined as 
\begin{align}
\label{covder}
 {\cal D}_\s =  \d_\s  - \frac{i}{2} n \left[ \PS_{\s \lambda} + {(1-\frac{3}{2}
\eL)} \PT_{\s \lambda} \right]  \g^\lambda   \ .
\end{align}

The mass $m$ is obtained from plugging  (\ref{Tperfect}) in (\ref{general mass}). This leads to a second degree equation. We arbitrarily choose the  root leading to a positive mass
\begin{align}
m  =  \frac{ \sqrt{ 3\eps }}{ 4 M_P} ~ |\frac{1}{3} - w| \ .
\end{align}
Notice that for $\eL = 0$ this expression is equal to the Hubble parameter of an FRW metric that would be generated by having $T^{\mu \nu}$ on the r.h.s of Einstein equations.

\section{Explicit decomposition of a spin-3/2 in helicity-operator eigenstates}
\label{Appdirectdecomposition}

We review in this section the explicit decomposition of a spin-3/2 particle in the case of a fluid at rest. We follow the Wess and B\"agger notations in considering
$ \eta^{\mu \nu} = \textrm{diag}(-1,+1,+1,+1)$ and the gamma matrices in the Weyl basis such that
\begin{align}
\g^0 = \begin{pmatrix}0 & \s^0 \\ \s^0 & 0 \end{pmatrix} & \, &  \g^0 = \begin{pmatrix}0 & \s^i \\ -\s^i & 0 \end{pmatrix}  &\, & \g^5 = \g^0 \g^1 \g^2 \g^3 = \begin{pmatrix} -i\mathbf{1} & 0 \\ 0& i\mathbf{1} \end{pmatrix} \ ,
\end{align}
with the anti-commutation relation $\left\{ \g^\mu , \g^\nu \right\} = -2 \eta^{\mu \nu}$. With these conventions $\tilde{S}^{\mu \nu}$ and $S^{\mu \nu}$ defined by
\begin{align}
\tilde{S}^{\mu \nu} = \frac{- i }{ 4 } (\g^\mu \g^\nu - \g^\nu \g^\mu) \ ,
\end{align}
and
\begin{align}
(S^{\mu \nu})_m^n = i (\delta^\mu_m \eta^{n \nu} - \delta^\nu_m \eta^{n \mu}) \ ,
\end{align}
form a representation of the Lorentz group on the spinor and vector space respectively. The rotation generators on the spinor-vector representation are the $J_i = \frac{1}{2} \epsilon_{i j k } (\tilde{S}^{j k} + S^{j k})$, where $\epsilon_{i j k }$ is the fully antisymmetric tensor such that $\epsilon_{123} = 1$.

In particular, considering a momentum $p^\mu = (p^0, \vec{k})$ with $p^2 = k^2 - (p^0)^2$, the helicity operator along $\vec{k}$ is
\begin{align}
\mathcal{S} ~\equiv~ \frac{k^i}{k} J_i = \frac{k^i}{k} ( \frac{1}{2} \tilde{\Sigma}_i + \Sigma_i) \ ,
\end{align}
where $k = \sqrt{k^2}$
\begin{align}
\tilde{\Sigma}_i = (S^{23}, S^{31} , S^{12}) \\[1.2em]
\Sigma_i =  \begin{pmatrix}  \s^i  & 0 \\ 0&\s^i \end{pmatrix} = i \g^5 \g^0 \g_i \ .
\end{align}
Eigenvectors of the helicity operator in the vector space are the $\epsilon_0'$ and $\epsilon_0, \epsilon_+, \epsilon_-$ corresponding to $j=0, h=0$ and $j=1, h=0, +1 , -1$ respectively (with helicity eigenvalues labeled by $h$ and those of $J^2$ by $j(j+1)$ ). They are easily obtained when $\vec{k} = (0,0,k)$ as
\begin{align}
&& && {\epsilon_0}^\mu &= (\frac{k}{|p|},0,0,\frac{p^0}{|p|}) \\
{\epsilon_0'}^\mu = \frac{p^\mu}{|p|} && \textrm{  and  } && {\epsilon_+}^\mu &= \frac{1}{\sqrt{2}} (0,-1,i,0) \\
 && &&{\epsilon_-}^\mu &= \frac{1}{\sqrt{2}} (0,1,i,0) \ ,
\end{align}
where we have taken $|p| = \sqrt{-p^2}$. Also note that they are normalised by : ${{\epsilon'_0}^{*}}^\mu {\epsilon'_0}_\mu = -1$ and ${\epsilon_{0,+,-}^{*}}^\mu {\epsilon_{0,+,-}}_\mu = 1$. For a general $\vec{k}$ obtained by rotating with $\theta$ around the $y$ axis and $-\phi$ around the $z$ axis
\begin{align}
p^\mu &= \left( p^0 ,~ k \cos \phi  \sin\theta  ,~ k \sin\phi \sin\theta ,~ k\cos\theta  \right) \ ,
\end{align}
helicity eigenvectors are given by\footnote{Since we have chosen the opposite signature from~\cite{Auvil:1966} and~\cite{Moroi:1995fs}, helicity +1 and -1 eigenvectors are inverse compared to those of these authors.}:
\begin{align}
\begin{split}
{\epsilon'_0}^\mu &= \frac{p^\mu}{|p|}  \\
{\epsilon_+}^\mu &= \frac{1}{\sqrt{2}} \left( 0, - \cos\theta\cos\phi - i \sin\phi, -\cos\theta \sin\phi + i \cos\phi, \sin\theta \right) \\
{\epsilon_-}^\mu &= \frac{1}{\sqrt{2}} \left(0,\cos\theta \cos\phi - i \sin\phi, \cos\theta \sin\phi + i \cos\phi, -\sin\theta \right) \\
{\epsilon_0}^\mu &= \left( \frac{k}{|p|},\frac{p^0}{|p|} \frac{\vec{k}}{k} \right) \ .
\end{split}
\end{align}

Finally, the spinor-vector representation of the Lorentz group (written as representation of $SU(2)_L \times SU(2)_R$ for a left-handed Weyl spinor) can be decomposed into spin representations as
\begin{align}
(\frac{1}{2},\frac{1}{2}) \otimes (\frac{1}{2},0) =  \frac{1}{2} \oplus (1 \otimes \frac{1}{2}) = \frac{1}{2} \oplus \frac{1}{2} \oplus \frac{3}{2} \ ,
\end{align}
In term of states, we can therefore decompose $\left| \psi \right\rangle$ as
\begin{align}
\begin{split}
\left| \psi \right\rangle = &a_1 \left|\frac{1}{2}, \frac{1}{2} \right\rangle' + \tilde{a}_1  \left|\frac{1}{2}, -~\frac{1}{2} \right\rangle' + a_2 \left|\frac{1}{2}, \frac{1}{2} \right\rangle + \tilde{a}_2 \left|\frac{1}{2}, -~\frac{1}{2} \right\rangle \\
& +  a_3 \left|\frac{3}{2}, \frac{1}{2} \right\rangle + \tilde{a}_3 \left|\frac{3}{2}, -~\frac{1}{2} \right\rangle + a_4 \left|\frac{3}{2}, \frac{3}{2} \right\rangle + \tilde{a}_4  \left|\frac{3}{2}, -~\frac{3}{2} \right\rangle \ ,
\end{split}
\end{align}
where the prime notes the first spin-1/2 representation. Using the Clebsch-Gordon decomposition this last expression gives~(\ref{explicitdecompo}).

If we boost from the fluid rest frame to any frame, the eigenvector $\epsilon_0$ becomes:
\begin{align}
\epsilon_0^\mu =  \frac{k}{|p|} u^\mu + \frac{q}{|p| k } k^\mu  \ ,
\end{align}
where we have $q^\mu = \PT^{\mu \nu} p_\nu$ and  $k^\mu = \PS^{\mu \nu} p_\nu$ with $q = \sqrt{-q^2}$. The form of $\epsilon_+$ and $\epsilon_-$ is complex in general, but we still have the property: $u_\mu \epsilon_{\pm}^\mu = 0 $ since $u_\mu \epsilon_{\pm}^\mu$ is a Lorentz invariant.

\section{Computation of the covariant propagator} 
\label{AppPropagator}
 We assume the approximation~(\ref{neglectder}) so that we can use plane wave solutions of momentum $p^\mu$.

In making calculations, it is helpful to   use the commutations or
anti-commutations properties of the operators $\pi_i$ with 
$\ks$, $\ps$ and $\qs$ which are (omitting the Lorentz index for clarity) given by:
\begin{align}
\label{commutation}
 \pi_1 \ks &=   \ks \pi_1 &   \pi_1 \qs &=   \qs \pi_1 \\
 \pi_2 \ks &=  -\ks \pi_2 &  \pi_2 \qs &=   -\qs \pi_2 \\
 \pi_3 \ks &=  -\ks \pi_3 &   \pi_3 \qs &=   -\qs \pi_3 \ .
\end{align}

The contraction rules of this set of projectors is straightforward and can be
summarised in Table 1.
\begin{table}
\label{algebraproj}
\scriptsize 
\begin{tabular}{|c | c c c c c c c c c|}
\hline
           &  $\P_{11}$ & $\P_{12}$ & $\P_{13}$ & $\P_{21}$ & $\P_{22}$ &
$\P_{23}$ & $\P_{31}$ & $\P_{32}$ & $\P_{33}$ \\
\hline 
 $\P_{11}$ &  $\P_{11}$ & $\P_{12}$ & $\P_{13}$ & $0$ & $0$ & $0$ & $0$ & $0$ &
$0$ \\[0.5em]
 $\P_{21}$ &  $\P_{21}$ & $\P_{22} / (\pi_1^2 \pi_2^2)$ & $\P_{23}/ (\pi_1^2)$ &
$0$ & $0$ & $0$ & $0$ & $0$ & $0$ \\[0.5em]
 $\P_{31}$ &  $\P_{31}$ & $\P_{32} / (\pi_1^2)$ & $\P_{33}/ (\pi_1^2 \pi_2^2)$ &
$0$ & $0$ & $0$ & $0$ & $0$ & $0$ \\[0.5em]
 $\P_{12}$ & $0$ & $0$ & $0$ &  $\P_{11} / (\pi_1^2 \pi_2^2)$ & $\P_{12} $ &
$\P_{13} / (\pi_2^2)$  & $0$ & $0$ & $0$ \\[0.5em]
 $\P_{22}$ & $0$ & $0$ & $0$&  $\P_{21}$ & $\P_{22} $ & $\P_{23}$  & $0$ & $0$ &
$0$ \\[0.5em]
 $\P_{32}$ & $0$ & $0$ & $0$&  $\P_{31}  / (\pi_2^2)$ & $\P_{32} $ & $\P_{33}/
(\pi_1^2 \pi_2^2)$  & $0$ & $0$ & $0$ \\[0.5em]
 $\P_{13}$ & $0$ & $0$ & $0$ & $0$ & $0$ & $0$&  $\P_{11} / (\pi_1^2 \pi_3^2)$ &
$\P_{12} / (\pi_3^2)$ & $\P_{13}$  \\[0.5em]
 $\P_{23}$ & $0$ & $0$ & $0$ & $0$ & $0$ & $0$&  $\P_{21}/ (\pi_3^2)$ & $\P_{22}
/ (\pi_3^2 \pi_2^2)$ & $\P_{23}$  \\[0.5em]
 $\P_{33}$ & $0$ & $0$ & $0$ & $0$ & $0$ & $0$&  $\P_{31}$ & $\P_{32}$ &
$\P_{33}$  \\[0.5em]
\hline
\end{tabular}
\caption{\small Contraction rules for the nine projectors $\P_{i,j}$, the
extra-factors of $\pi_i^2$ comes from the normalisation of the nilpotent
operators.}
\end{table}
We supplemented this set of projector by $\P_{\thalf}$ defined such that:
\begin{equation}
 \P_{\thalf}^{\mu \nu} + \P_{33}^{\mu \nu} +\P_{22}^{\mu \nu} +\P_{11}^{\mu \nu} =
\eta^{\mu \nu} \ .
\end{equation}

We use the identities
\begin{align}
 p^\mu  &=   \pi_1^\mu  &  \g^\mu &=   \frac{\ps}{p^2} \pi_1^\mu -
\frac{\ps}{k^2} \pi_2^\mu + \pi_3^\mu \\
k^\mu &=  \frac{k^2}{p^2} \pi_1^\mu + \frac{\ks \qs}{k^2} \pi_2^\mu& r^\mu &=  
\frac{\ks}{p^2} \pi_1^\mu - \frac{\qs}{k^2} \pi_2^\mu + \pi_3^\mu \\
q^\mu &= \frac{q^2}{p^2} \pi_1^\mu - \frac{\ks \qs}{k^2} \pi_2^\mu  & t^\mu &=  
\frac{\qs}{p^2} \pi_1^\mu - \frac{\ks}{k^2} \pi_2^\mu  \ ,
\end{align}
and a bit algebra to express $K^{\mu \nu}$ as a function of the $\P_{ij}^{\mu
\nu}$. We can write 
\begin{align}
 K^{\mu \nu} = (\ps + m ) (\P_{3/2}^{\mu \nu} - \P_{3 3}^{\mu \nu}) - U
(\P_{13}^{\mu \nu} - \P_{31}^{\mu \nu}) + V (\P_{23}^{\mu \nu} - \P_{32}^{\mu
\nu}) - W (\P_{12}^{\mu \nu} - 
\P_{21}^{\mu \nu}) \ ,
\end{align}
with
\begin{align}
&W = n k^2 \\[0.8em] 
&U = 2 (\ks m + \qs n) \\
&V = 2 \frac{k^2}{p^2} (p^2 - m \qs - n\ks)  \ .  
\end{align}
The second calculation trick is to define a conjugation relation by 
\begin{align}
 \overline{A} =  \overline{A_1 + \ks A_2 + \qs A_3} = A_1 - \ks A_2 - \qs A_3 \
,
\end{align}
this operation satisfies all the usual properties of conjugation 
$\overline{AB} = \overline{A} ~ \overline{B} $, $\overline{A+B} = \overline{A} +
\overline{B} $, and we have also 
\begin{align}
|A|^2 =  A \overline{A} =  a_1^2 + k^2 a_2^2 + q^2 a_3^3 \ ,
\end{align}
which enables us to obtain the inverse as 
\begin{align}
\label{inverse}
 A^{-1}  =  \frac{\overline{A}}{A \overline{A}} = \frac{A_1 - \ks A_2 - \qs
A_3}{ a_1^2 + k^2 a_2^2 + q^2 a_3^3} \ .
\end{align}

This formula uses the assumption that all parameters
$a_i$ are scalars. If $A$ is such that $a_1 = a_1' +
\ks \qs a_1''$ 
the previous formula makes little sense. However, it is always
possible in this case to factorise $A = A_1 A_2$ where $A_1$ and $A_2$ have only
scalar 
coefficients.\footnote{If we had $C = c_1 + c_1' \ks\qs + c_2 \ks + c_3 \qs$,
one can easily check that $C = (c_3 + c_1' \ks + (\frac{c_2 c_3}{c_1'}- c_1)
\frac{1}{q^2}\qs)(\frac
{c_2}{c_1'} + \qs)  $ is such a decomposition.}

A crucial observation is that thanks to the relations~(\ref{commutation}), the
(anti)commutation relations between the operators $\pi_i$ and $A$ are
\begin{align} 
\begin{aligned}
 \pi_1^\mu A &=   A \pi_1^\mu   \\
\pi_2^\mu A &=  \overline{A} \pi_2^\mu  \\
 \pi_3^\mu A &=  \overline{A} \pi_3^\mu  \  ,
\end{aligned}
\end{align}
which allows to make all  calculations using the $A$ form without
decomposing it in $\ks$ or $\qs$ parts. The decomposition of the propagator on the
projectors basis can be written as
\begin{align}
\begin{split}
G^{\mu \nu} = \frac{ m  - \ps}{m^2 + p^2} \P_{\thalf}^{\mu \nu} + & A ~ \P_{1
1}^{\mu \nu} + B ~ \P_{2 2}^{\mu \nu} +  C ~\P_{3 3}^{\mu \nu} + D ~\P_{13}^{\mu
\nu}  + D' ~\P_{31}^
{\mu \nu}  \\ 
& + E ~\P_{23}^{\mu \nu} + E' ~\P_{32}^{\mu \nu} + F ~\P_{12}^{\mu \nu}  + F'
~\P_{21}^{\mu \nu} \ .
\end{split}
\end{align}
The propagator satisfies
\begin{equation}
 K_{\mu \rho} G^{\rho \nu} = \eta_\mu^{~\nu} \ .
\end{equation}
Expanded it in the projectors basis leads to a system of nine equations
\begin{align} 
\begin{aligned}
 (33) & \quad &&- (\ps + m) C - (\pi_2 \pi_3 )^{-2} V E + (\pi_1 \pi_3 )^{-2} U
\overline{D} =  1   \\
 (32) & \quad &&- (\ps + m) E' - (\pi_1 )^{-2} U \overline{F} - V B =  0   \\
 (31) & \quad &&- (\ps + m) D' - (\pi_2 )^{-2} V F' +  U \overline{A} =  0   \\
 (23) & \quad &&   (\pi_1 )^{-2} W \overline{D} + V C =  0   \\
 (22) & \quad &&   (\pi_2 \pi_3 )^{-2} V E' + (\pi_1 \pi_2 )^{-2} W \overline{F}
=  1   \\
 (21) & \quad &&   (\pi_3 )^{-2} V D' + W \overline{A} =  0   \\
 (13) & \quad &&   -(\pi_2 )^{-2} W \overline{E} - U \overline{C} =  0   \\
 (12) & \quad &&    -(\pi_3 )^{-2} U \overline{E}' - W \overline{B} =  0   \\
 (11) & \quad &&    -(\pi_1 \pi_3 )^{-2} U \overline{D}' - (\pi_1 \pi_2 )^{-2} W
\overline{F}' =  1    \ .
\end{aligned}
\end{align}
We  make the assumption that $U, V$ and  $W$ contains no terms in $\ks \qs$
( if it was not the case, one should first factorised it, then inverse both
terms using~(\ref
{inverse}) and do the same to inverse $X$). We define the quantity $X$ by 
\begin{align}
 X = -(\ps + m) + \frac{1}{\pi^3 W}(V W \overline{U} - U \overline{W} V) \ .
\end{align}
This expression has no terms in $\ks \qs$. Indeed, if $A$ and $B$ do
not have $\ks \qs$ terms, then it is easily seen that $AB + BA$  also does not
have $\ks \qs$ 
terms and that $AB + \overline{BA}$ is a pure scalar using the product formula 
\begin{align}
 (a_1 + a_2 \ks + a_3 \qs)(b_1 + b_2 \ks + b_3 \qs) = &~ (a_1 b_1 - a_2 b_2 k^2
- a_3 b_3 q^2) + (a_2 b_1 + a_1 b_2) \ks \nonumber \\
 & + (a_3 b_1 + a_1 b_3) \qs + (a_2 b_3 - b_2 a_3) \ks \qs \ .
\end{align}
Since we can write
\begin{equation}
 V W \ov{U} - U \ov{W} V = V (W \ov{U}) + (W \ov{U}) V  - (W \ov{U} + U \ov{W})
V  \ ,
\end{equation}
we conclude that one can apply the formula~(\ref{inverse}) on $X$ and solve the
system of equations as
\begin{align} 
\begin{aligned}
A &= \frac{\pi_1^2}{\pi_3^2} \frac{W}{|W|^2} \ov{V} \frac{X}{|X|^2}  \ov{V}
\frac{\ov{W}}{|W|^2} && \\
B &= \frac{\pi_2^2}{\pi_3^2} \frac{W}{|W|^2} \ov{U} \frac{\ov{X}}{|X|^2} U
\frac{\ov{W}}{|W|^2} && \\
C &= \frac{\ov{X}}{|X|^2} && \\
D &= -\pi_1^2  \frac{W}{|W|^2} \ov{V} \frac{X}{|X|^2}  &  D' &= -\pi_1^2 
\frac{\ov{X}}{|X|^2} V \frac{W}{|W|^2}  \\
E &= -\pi_2^2  \frac{W}{|W|^2} \ov{U} \frac{\ov{X}}{|X|^2}  &  E' &= -\pi_2^2 
\frac{\ov{X}}{|X|^2} U \frac{\ov{W}}{|W|^2}  \\
F &= \pi_1^2 \pi_2^2 (\frac{W}{|W|^2}  + \frac{1}{\pi_3^2}\frac{W}{|W|^2} \ov{V}
\frac{X}{|X|^2}  \ov{U} \frac{W}{|W|^2} )&& \\
F' &= \pi_1^2 \pi_2^2 ( - \frac{W}{|W|^2}  + \frac{1}{\pi_3^2}\frac{W}{|W|^2}
\ov{U} \frac{\ov{X}}{|X|^2}  V \frac{W}{|W|^2} )&& \
\end{aligned}
\end{align}
Replacing these expression in the propagator, we can in fact factorise most of
these terms and obtain:
\begin{align}
\begin{aligned}
 G^{\mu \nu} = \frac{m - \ps}{p^2 + m^2}\P_{\thalf}^{\mu \nu} &+ \frac{1}{\pi_3^2
} \left[\frac{W}{|W|^2} \ov{V} \pi_1^\mu + \frac{W}{|W|^2} \ov{U} \pi_2^\mu -
\pi_3^\mu \right] \frac{X}{|X|^2} 
\left[\ov{V} \frac{\ov{W}}{|W|^2}  \pi_1^\nu + \ov{U} \frac{W}{|W|^2}  \pi_2^\nu -
\pi_3^\nu \right]   \\
& + \frac{W}{|W|^2}(\pi_1^\mu \pi_2^\nu - \pi_2^\mu \pi_1^\nu) \ .
\end{aligned}
\end{align}
If we replace  $U,V $ and $W$ by their expression in our case we observe first
that $X$ simplifies in 
\begin{align}
 X = 3 (m - w\ks + \qs) \ ,
\end{align}
which indeed does not include $\ks \qs$ terms. Replacing in the propagator, we
recover the expression used previously 
\begin{align}
 G^{\mu \nu} &= \frac{\Pi_{\thalf}^{\mu \nu}}{p^2 + m^2} + \frac{\Pi_{\half}^{\mu
\nu}}{m^2 + w^2 k^2 + q^2}  + \frac{3}{4} \eL \frac{\ks}{m k^2}(t^\mu p^\nu - 
p^\mu t^\nu) \ .
\end{align}
where the two polarisations can be written 
\begin{align}
\Pi_{\thalf}^{\mu \nu} = & (m - \ps) \P_{\thalf}^{\mu \nu} 
\end{align}
and
\begin{align}
\begin{split}
\Pi_{\half}^{\mu \nu} =&  - \frac{2}{3} \    \Lambda^\mu \   (\ps +  m - \eL \ks )\  \Lambda^\nu  \ .
\end{split} 
\end{align}
where
\begin{equation}
\Lambda^\mu = \g^\mu + \frac{p^\mu}{n} - \frac{3}{2} (r^\mu - \frac{\slashed k k^\mu}{k^2}) - \frac{3}{4} \eL t^\mu  \ .
\end{equation}

We can relate $\Pi_{\thalf}^{\mu \nu}$ to the Rarita-Schwinger polarisation tensor $\Pi_{RS}^{\mu \nu}$ defined as
\begin{align}
\Pi_{RS}^{\mu \nu} = (m - \ps) [\eta^{\mu \nu} + \frac{1}{3} \g^\mu \g^\nu + 2
\frac{p^\mu p^\nu}{3m^2} + \frac{\g^\mu p^\nu - \g^\nu p^\mu}{3 m} ]\ ,
\end{align}
by
\begin{align}
 \P_{\thalf}^{\mu \nu} = \Pi_{RS}^{\mu \nu} + \frac{2}{3} [\Lambda^\mu + \frac{3}{4} \eL (t^\mu + \frac{p^\mu}{m})](\ps + m) [\Lambda^\nu + \frac{3}{4} \eL (t^\nu + \frac{p^\nu}{m})] \ .
\end{align}
This expression makes explicit the fact that our propagator reduces to the usual Rarita-Schwinger result when $\eL = 0$. It is remarkable that $\Pi_{\thalf}^{\mu \nu}$ is directly related to $\P_{\thalf}^{\mu \nu} $ since this implies that the helicity-3/2 components have a pole structure of a relativistic particle corresponding to their origins as the gravitino. Similarly, albeit the expression of $\Pi_{\half}^{\mu \nu}$ looks different from $P_{\half}^{\mu \nu}$, an explicit calculation using constraint~(\ref{C1}) shows that  we have :
\begin{align}
 \Pi_{\half}^{\mu \nu} \psi_\nu = \Pi_{\half}^{\mu \nu} {\psih}_\nu \ .
\end{align}
Therefore, the helicity-1/2 part of the propagator projects naturally on the correct helicity-1/2 degrees of freedom, which means that they have the pole structure of a spin-1/2 pseudo-particle of velocity $w$.

\end{document}